\documentclass[11pt]{article}

\usepackage{fullpage}
\usepackage{microtype}
\usepackage{epsfig}
\usepackage{microtype}
\usepackage{graphicx}
\usepackage[linesnumbered,ruled, lined, noend]{algorithm2e}
\usepackage{algpseudocode}
\usepackage{epsfig,enumerate,amsmath,amsfonts,amssymb,amsthm,mathrsfs,ifpdf}
\usepackage{indentfirst,relsize}
\usepackage[numbers]{natbib}
\usepackage{setspace}
\usepackage{enumerate}
\usepackage{latexsym}
\usepackage{stackrel}
\usepackage[all]{xy}
\usepackage[usenames,dvipsnames]{pstricks}
\usepackage{pst-grad}
\usepackage{pst-plot}
\usepackage{xspace}
\usepackage{hyperref}

\newcommand{\size}[1]{\left| #1 \right|}

\newcommand{\remove}[1]{}

\newcommand{\R}{\mathbb{R}}
\newcommand{\N}{\mathbb{N}}
\newcommand{\Z}{\mathbb{Z}}
\newcommand{\cI}{\mathcal{I}}

\newcommand{\cS}{\mathcal{S}}
\newcommand{\cT}{\mathcal{T}}

\newcommand{\cL}{\mathcal{L}}

\newcommand{\cP}{\mathcal{P}}

\newcommand{\cR}{\mathcal{R}}
\newcommand{\cG}{\mathcal{G}}

\theoremstyle{plain}
\newtheorem{theo}{Theorem}

\newtheorem{pro}[theo]{Proposition}

\newtheorem{conj}{Conjecture}
\newtheorem{cl}[theo]{Claim}

\theoremstyle{definition}

\newtheorem{rem}{Remark}
\newtheorem{obs}[theo]{Observation}

\newtheorem{problem}[theo]{Problem}

\newcommand{\disj}[1]{\mbox{{\sc Disj}}_{#1}}
\newcommand{\intsec}[1]{\mbox{{\sc Int}}_{#1}}

\newcommand{\spdisj}[2]{#2\mbox{{\sc -SparseDisj}}_{#1}}
\newcommand{\spint}[2]{#2\mbox{{\sc -SparseInt}}_{#1}}

\newcommand{\augind}[1]{$\mbox{{\sc AugIndex}}_{#1}$}

\newcommand{\val}[1]{{\textit{decimal}}({#1})}

\newcommand{\disjline}{{\sc Discrete Line Disj}\xspace}
\newcommand{\intline}{{\sc Discrete Line Int}\xspace}
\newcommand{\disjinter}{{\sc Discrete Interval Disj}\xspace}

\newcommand{\defproblem}[3]{
  \vspace{1mm}
\noindent\fbox{
  \begin{minipage}{0.96\textwidth}
  \begin{tabular*}{\textwidth}{@{\extracolsep{\fill}}lr} #1 \\ \end{tabular*}
  {\bf{Input:}} #2  \\
  {\bf{Output:}} #3
  \end{minipage}
  }
  \vspace{1mm}
}

\title{
Disjointness through the Lens of Vapnik–Chervonenkis\\ 
Dimension: Sparsity and Beyond
}

\author{
Anup Bhattacharya\footnote{Indian Statistical Institute, Kolkata, India}
\and 
Sourav Chakraborty\footnotemark[1]
\and 
Arijit Ghosh\footnotemark[1]
\and 
Gopinath Mishra\footnotemark[1]
\and 
Manaswi Paraashar\footnotemark[1]
}

\date{}

\begin{document}

\maketitle
\thispagestyle{empty} 

\begin{abstract}
The disjointness problem - where Alice and Bob are given two subsets of $\{1, \dots, n\}$ and they have to check if their sets intersect - is a central problem in the world of communication complexity. While both deterministic and randomized communication complexities for this problem are known to be $\Theta(n)$, it is also known that if the sets are assumed to be drawn from some restricted set systems then the communication complexity can be much lower. In this work we explore how communication complexity measures change with respect to the complexity of the underlying set system. The complexity measure for the set system that we use in this work is the Vapnik-Chervonenkis (VC) dimension. More precisely, on any set system with VC dimension bounded by $d$, we analyze how large can the deterministic and randomized communication complexities be, as a function of $d$ and $n$.  The $d$-sparse set disjointness problem, where the sets have size at most $d$, is one such set system with VC dimension $d$. The deterministic and the randomized communication complexities of the $d$-sparse set disjointness problem have been well studied and is known to be $\Theta \left( d \log \left({n}/{d}\right)\right)$ and $\Theta(d)$, respectively, in the multi-round communication setting. In this paper, we address the question of whether the randomized communication complexity is always upper bounded by a function of the VC dimension of the set system, and does there always exist a gap between the deterministic and randomized communication complexity for set systems with small VC dimension.

In this paper, we construct two natural set systems of VC dimension $d$, motivated from geometry. Using these set systems we show that the deterministic and randomized communication complexity can be $\widetilde{\Theta}\left(d\log \left( n/d \right)\right)$ for set systems of VC dimension $d$ and this matches the deterministic upper bound for all set systems of VC dimension $d$. We also study the deterministic and randomized communication complexities of the set intersection problem when sets belong to a set system of bounded VC dimension. We show that there exists set systems of VC dimension $d$ such that both deterministic and randomized (one-way and multi-round) complexity for the set intersection problem can be as high as $\Theta\left( d\log \left( n/d \right) \right)$, and this is tight among all set systems of VC dimension $d$.

\end{abstract}

\newpage
\pagenumbering{arabic}

\newpage

\section{Introduction}

Since its introduction by Yao~\cite{Yao79}, communication complexity occupies a central position in theoretical computer science. A striking feature of communication complexity is its interplay with other diverse areas like analysis, combinatorics, and geometry~\cite{10.5555/264772,DBLP:journals/fttcs/Roughgarden16}. Vapnik and Chervonenkis~\cite{VC71} introduced the measure {\em Vapnik-Chervonenkis dimension} or {\em VC dimension} for set systems in the context of statistical learning theory. As was the case with communication complexity, VC dimension has found numerous connections and applications in many different areas like approximation algorithms, discrete and combinatorial geometry, computational geometry, discrepancy theory and many other areas~\cite{matousek2009geometric,chazelle2001discrepancy,pach2011combinatorial,matousek2013lectures}.
In this work we study both of them under the same lens: of restricted systems and, for the first time, prove that geometric simplicity does not necessarily imply efficient communication complexity.

Lets start with recollecting some definitions from Vapnik–Chervonenkis theory. Let $\mathcal{S}$ be a collection of subsets of a {\em universe}  $\mathcal{U}$.
For a subset $y$ of $\mathcal{U}$, we define 
$$ 
    \mathcal{S}{{\mid_{y}}} := \left\{ y \cap x \, : \, x \in \mathcal{S}\right\}.
$$
We say a subset $y$ of $\mathcal{U}$ is {\em shattered} by $\mathcal{S}$ if $\mathcal{S}{{\mid_{y}}} = 2^{y}$, where $2^{y}$ denotes the power set of $y$. {\em Vapnik–Chervonenkis (VC) dimension} of $\mathcal{S}$, denoted as $\mbox{VC-dim}(\mathcal{S})$, is the size of the largest subset $y$ of $\mathcal{U}$ shattered by $\mathcal{S}$. 
VC dimension has been one of the fundamental measures for quantifying
complexity of a collection of subsets. 


Now let us revisit the world of communication complexity. Let $f: \Omega_{1} \times \Omega_{2} \rightarrow \Omega$. In
{\em communication complexity}, two players Alice and Bob get as inputs $x
\in \Omega_{1}$  and $y \in \Omega_{2}$ respectively, and the goal for
the players is to device a protocol to compute $f(x,y)$ by exchanging 
as few bits of information
between themselves as
possible. 

\remove{Since its introduction by Yao~\cite{Yao79}, communication
complexity has found many applications in different areas of computer
science like streaming algorithms, property testing, sketching, data
structures, circuit complexity and auction theory~\cite{10.5555/264772,DBLP:journals/fttcs/Roughgarden16}. \textcolor{blue}{Manaswi: this is a repetition of para 1. Also, we should add some references here and later. We are citing the three books quite a few time.}
}

The {\em deterministic communication complexity} $D(f)$ of a function $f$
is the minimum number of bits Alice and Bob will exchange in the worst case
to deterministically compute the function $f$. In the randomized setting, both Alice and Bob
share an infinite random source\footnote{This is the communication complexity setting with shared random coins. If no random string is shared, it is called the private random coins setting. By \cite{newman} we know that the communication complexity in both the setting differs by at most a logarithmic additive factor.}
and the goal is to give the correct answer with probability at
least $2/3$. The randomized
communication complexity $R(f)$ of $f$ denotes the minimum number of
bits exchanged by the players in the worst case input by the best randomized
protocol computing $f$. In both deterministic and randomized settings, Alice and Bob are 
allowed to make multiple rounds of interaction. Communication complexity when the number of rounds of interaction 
is bounded is also often studied. An important special case is when only one round of communication is allowed,  
that is, only Alice is allowed to send messages to Bob and Bob computes the output. 
We will denote by $D^{\rightarrow}(f)$  and $R^{\rightarrow}(f)$ the {\em one way deterministic communication complexity} and {\em one way randomized communication complexity} respectively, 
of $f$.




One of the most well studied functions in communication complexity is the disjointness function.  Given a universe $\mathcal{U}$ known to both Alice and Bob, the
{\em disjointness function}, $\disj{\mathcal{U}}: 2^{\mathcal{U}} \times 2^{\mathcal{U}}
\rightarrow \{0,\, 1\}$, where
$2^{\mathcal{U}}$ denotes the power set of $\mathcal{U}$, is defined as follows:
\begin{equation}
  \disj{\mathcal{U}}(x,y) =
  \begin{cases}
    1, & \text{if } x \cap y = \emptyset \\
    0, & \text{o/w }
  \end{cases}
\end{equation}
We also define the {\em intersection function}. Given a universe $\mathcal{U}$ known to both Alice and Bob, the
{\em intersection function}, $\intsec{\mathcal{U}}: 2^{\mathcal{U}} \times 2^{\mathcal{U}}
\rightarrow 2^{\mathcal{U}}$, where
$2^{\mathcal{U}}$ denotes the power set of $\mathcal{U}$, is defined as
$\intsec{\mathcal{U}}(x,y) = x \cap y$.
It is easy to see that $\intsec{\mathcal{U}}$ is harder function to compute than $\disj{\mathcal{U}}$.
\remove{
Also defining $t$-sparse set disjointness: Given a universe $\mathcal{U}$ known to both Alice and Bob and parameter $t \in \N$. Let $\mathcal{S}|_t$ be the collection of all subsets of $\mathcal{U}$ of cardinality at most $t$, the
{\em sparse set disjointness},
$\spdisj{\mathcal{U}}{t}: \mathcal{S}|_t \times \mathcal{S}|_t
\rightarrow \{0,1\}$ for all $x,y \in \mathcal{S}|_t$, 
\begin{equation}
  \intsec{\mathcal{U}}(x,y) =
  \begin{cases}
    1, & \text{if } x \cap y = \emptyset \\
    0, & \text{o/w }
  \end{cases}
\end{equation}
}
The $\disj{\mathcal{U}}$ function and its different variants, like $\intsec{\mathcal{U}}$, have been one of the most important problems
in communication complexity and have found numerous applications in
areas like streaming algorithms for proving lower 
bounds~\cite{DBLP:journals/fttcs/Roughgarden16, rao_yehudayoff_2020}. By abuse of the notation, when $\mathcal{U} = [n]$, where $[n]$ denotes the set $\{1, \, \dots, \, n\}$, we will denote the functions $\disj{[n]}$ and $\intsec{[n]}$ by $\disj{n}$
and $\intsec{n}$ respectively.


Using the standard {\em rank argument}~\cite{10.5555/264772,rao_yehudayoff_2020} one can
show that $D(\disj{n}) = \Theta(n)$. 
In a breakthrough paper, Kalyanasundaram and Schnitger~\cite{KalyanasundaramS92} proved that $R(\disj{n}) = \Omega(n)$.
Razborov~\cite{Razborov92} and Bar-Yossef et al.~\cite{Bar-YossefJKS04} gave alternate proofs
for the above result. From the above cited results we can also see the $D(\intsec{n}) = R(\intsec{n}) = \Theta(n)$. 


Naturally, one would also like to ask what happens to the deterministic and randomized communication complexity (one way or multiple round) of $\disj{n}$, when both Alice and Bob know that their inputs have more structure. In particular what can we say if the inputs are guaranteed to be from a subset of $\mathcal{S}\subseteq 2^{\mathcal{U}}$, where $\mathcal{S}$ is known to both players.
Let $\disj{\mathcal{U}}$  functions {\em restricted} to $\mathcal{S} \times \mathcal{S}$ be denoted
by $\disj{\mathcal{U}}\mid_{\mathcal{S} \times \mathcal{S}}$. 
This problem has also been studied extensively, mostly for certain special classes of subsets $\mathcal{S}\subseteq 2^{\mathcal{U}}$. For example, the {\em sparse set disjointness} function, where the set $\mathcal{S}$ contains all the subsets of $\mathcal{U}$ of size at most $d$, is an important special case of these works.  

We will denote by $\spdisj{n}{d}$ and $\spint{n}{d}$, the functions $\disj{n}\mid_{S \times \mathcal{S}}$ and $\intsec{n}\mid_{S \times S}$ respectively, where $S$ is the collections of all subsets of $[n]$ of size at most $d$. 
Using the {\em rank argument}~\cite{10.5555/264772,rao_yehudayoff_2020}, one can again show
that, for all $d \leq n$, the deterministic communication complexity of
$\spdisj{n}{d}$  is $\Omega\left( d \log \frac{n}{d} \right)$.
H{\aa}stad and Wigderson~\cite{HastadW07}, and Dasgupta et
al.~\cite{DasguptaKS12} showed that the randomized 
communication complexity and one round randomized communication complexity of 
$\spdisj{n}{d}$ is  $R(\spdisj{n}{d}) = \Theta(d)$ and $R^{\rightarrow}(\spdisj{n}{d}) = \Theta(d \log d)$ respectively.
In a follow up work, Saglam and Tardos~\cite{SaglamT13} proved that with
$O(\log^{*}d)$ rounds of communication and $O(d)$ bits of
communication it is possible to compute $\spdisj{n}{d}$. 
More recently, Brody et al.~\cite{BrodyCKWY14} proved that 
$R^{\rightarrow} \left( \spint{n}{d} \right) = \Theta\left( d \log d\right)$ 
and $R(\spint{n}{d}) = \Theta(d)$. 
These results show that in the $d$-sparse setting, there is a separation between randomized and deterministic communication complexity of $\disj{n}$ and $\intsec{n}$ functions. 

One would like to ask what happens to the communication complexity for other restrictions to the disjointness (and intersection) problem. The following are two natural problems, with a geometric flavor, for which one would like to study the communication complexity.

\begin{problem}[\disjline]\label{problem1}
Let $G \subset \Z^2$ be a set of $n$ points in $\Z^2$ and $L$ be the set of all lines in $\R^2$. Also, let $\cL=L^d$ denote the collection of all $d$-size subsets of $L$.  The \disjline on $G$ and $\cL$ is a function, $ \disj{G}\mid_{\cL \times \cL}:\cL \times \cL \to \{0,1\}$ defined as $\disj{G}\mid_{\cL \times \cL} \left(\{\ell_1, \dots, \ell_d\}, \{\ell'_1, \dots, \ell'_d\}\right)$ is $1$ if and only if there exists a line in Alice's set\footnote{We assume that Alice has the set $\{\ell_1, \dots, \ell_d\}$ and Bob has the set $\{\ell'_1, \dots, \ell'_d\}$.} that intersects some line in Bob's set
at some point in $G$. Formally, 
\begin{equation}
 \disj{G}\mid_{\cL \times \cL}\left(\{\ell_1, \dots, \ell_d\}, \{\ell'_1, \dots, \ell'_d\}\right) =
  \begin{cases}
    1, & \text{if } \exists i, j \in [d]  \text{ s.t. } \ell_i \cap \ell'_j  \cap G =\emptyset \\
    0, & \text{o/w }
  \end{cases}
\end{equation}
\end{problem}

\begin{problem}[\disjinter]
Let $X \subset \Z$ be a set of $n$ points in $\Z$ and $Int$ be the set of all possible intervals. Also, let $\cI = Int^d$ denote the collection of all $d$-size subsets of $Int$. The \disjinter on $X$ and $\cI$ is a function, $\disj{X}\mid_{\cI \times \cI}:\cI \times \cI \to \{0,1\}$ defined as $\disj{X}\mid_{\cI \times \cI}\left(\{I_1,\, \dots,\, I_d\}, \{ I'_1, \, \dots, \, I'_d \} \right)$ is $1$ if and only if there exists an interval in Alice's set\footnote{We assume that Alice has the set $\{I_1, \dots, I_d\}$ and Bob has the set $\{I'_1, \dots, I'_d\}$.} that intersects some interval in Bob's set at some point in $X$.
\begin{equation}
  \disj{X}\mid_{\cI \times \cI}\left(\{I_1, \, \dots, \, I_d\}, \{I'_1, \, \dots, \, I'_d\}\right) =
  \begin{cases}
    1, &  \text{if } \exists i, j \in [d]  \text{ s.t. } I_i \cap I'_j \cap X = \emptyset \\
    0, & \text{o/w }
  \end{cases}
\end{equation}
\label{problem2}
\end{problem}


Note that both the \disjline and \disjinter functions are generalizations of sparse set disjointness function.\footnote{Take $n$ integer points on the $x$-axis. For \disjline setting, restrict only to lines orthogonal to $x$-axis. For \disjinter setting, take $n$ integer points on $\mathbb{Z}$ and only restrict to intervals containing one integer point. Both of these restriction will give the disjointness problem in the $d$-sparse setting.} 
Although it may not be obvious at first look, but both the \disjline function and the \disjinter functions  are disjointness functions restricted to a suitable subset.
In fact, the  connection between the {\em Sparse set disjointness} function ($\spdisj{n}{d}$), the  \disjline  function and the \disjinter function run deep - all the three subsets of the domain which help to define the functions as restriction of the disjointness function have VC dimension $\Theta(d)$, see Appendix~\ref{appendix-VCD}.
Naturally one would like to know, if the fact that the collection of subsets $\mathcal{S}$ has VC dimension $d$ has any implication on the communication complexity of $\disj{\mathcal{U}}\mid_{\mathcal{S} \times \mathcal{S}}$.
For example, is the randomized communication complexity of \disjline function and the \disjinter function upper bounded by a function of $d$ (independent of $n$)? And, do the \disjline function and the \disjinter function also have a separation between their randomized and deterministic communication complexities similar to that of the {\em Sparse set disjointness} function ($\spdisj{n}{d}$)? We show that these are not necessarily the cases.
\begin{theo}\label{thm:10way}
For \disjinter : there exists a $X \subset \Z$ with $n$ points such that    
      $$
      D(\disj{X}\mid_{\cI \times \cI}) = D^{\rightarrow}(\disj{X}\mid_{\cI \times \cI}) = R^{\rightarrow}(\disj{X}\mid_{\cI \times \cI}) = \Theta\left( d \log \frac{n}{d} \right).
      $$   
\end{theo}

\begin{theo}\label{thm:10way-1}
    For \disjline : there exists a $G \subset \Z^2$ with $n$ points such that $D(\disj{G}\mid_{\cL \times \cL})  = D^{\rightarrow}(\disj{G}\mid_{\cL \times \cL})= \Theta\left( d \log \frac{n}{d} \right)$ and, for the randomized setting, 
    $$
    R(\disj{G}\mid_{\cL \times \cL}) = \Omega\left( d \frac{\log (n/d)}{ \log \log (n/d)} \right)
    $$
\end{theo}

\remove{
A related function, to $\disj{\mathcal{U}}$, is the {\em Intersection} function  $\intsec{\mathcal{U}}: 2^{\mathcal{U}} \times 2^{\mathcal{U}} \rightarrow
2^{\mathcal{U}}$. It is defined as $\intsec{\mathcal{U}}(x\cap y) = x \cap y$. Note that $\intsec{\mathcal{U}}$ function is harder to compute than $\disj{\mathcal{U}}$ function. And just like is the case of $\disj{\mathcal{U}}$,
the $\intsec{\mathcal{U}}$  function {\em restricted} to $\mathcal{S} \times \mathcal{S}$ is denoted
by $\intsec{\mathcal{U}}\mid_{\mathcal{S} \times \mathcal{S}}$.


While $D(\spint{n}{d}) = \Theta(d\log \frac{n}{d})$, more recently, Brody et al.~\cite{BrodyCKWY14} proved that $R^{\rightarrow} \left( \spint{n}{d} \right) = \Theta\left( d \log d\right)$ and $R(\spint{n}{d}) = \Theta(d)$. These results show that behavior of $\disj{n}$ and $\intsec{n}$ functions, in the sparse setting is exactly same. }
\intline, that is, the intersection finding version of \disjline is defined as follows : the objective is to compute a function 
$\intsec {G}\mid_{\cL \times \cL}:\cL\times \cL \rightarrow G$ that is defined as 
$$
    \intsec{G}\mid_{\cL \times \cL}(\{\ell_1,\ldots,\ell_d\},\{\ell'_1,\ldots,\ell'_d\})=\bigcup\limits_{i,j \in [d]}\left(\ell_i \cap \ell'_j \cap G\right).
    \footnote{Agian, we will assume that Alice has the set $\{\ell_1, \dots, \ell_d\}$ and Bob has the set $\{\ell'_1, \dots, \ell'_d\}$.}
$$
As we have already mentioned,
Brody et al.~\cite{BrodyCKWY14} proved that $R(\spint{n}{d}) = \Theta(d)$, whereas 
$D(\spint{n}{d})=\Theta\left(d \log \frac{n}{d}\right)$. We show that 
\intline  does not demonstrate such a separation between the deterministic and randomized communication complexity.

\begin{theo}\label{thm:10way-2}
    For \intline : there exists a $G \subset \Z^2$ with $n$ points such that   $$
      D(\intsec{G}\mid_{\cL \times \cL}) = D^{\rightarrow}(\intsec{G}\mid_{\cL \times \cL}) =
      R^{\rightarrow}(\intsec{G}\mid_{\cL \times \cL})=
      R(\intsec{G}\mid_{\cL \times \cL}) = \Theta\left( d \log \frac{n}{d} \right).
      $$ 
\end{theo}

The upper bound for all the above three theorems can be obtained from the fact that the corresponding sets have VC dimension $\Theta(d)$, see Appendix~\ref{appendix-VCD}.
{\em Sauer-Shelah Lemma}~\cite{DBLP:journals/jct/Sauer72,S-Shelah72,VC71} 
		states that if $\mathcal{S} \subseteq 2^{[n]}$ and $\mbox{VC-dim}(\mathcal{S}) \leq d$ then 
		$\mid \mathcal{S}\mid \leq \left( \frac{e n}{d}
            \right)^{d}$. Thus if $\mbox{VC-dim}(\mathcal{S}) \leq d$,
		then  the Sauer-Shelah Lemma implies that 
		$D^{\rightarrow}(\intsec{n}\mid_{\mathcal{S}\times \mathcal{S}}) = O\left( d \log \frac{n}{d} \right)$. So, $O\left( d \log \frac{n}{d} \right)$ is a upper bound to the above questions, both for randomized and deterministic and also for the one-way communication.  But can the randomized communication complexity of $\disj{\mathcal{U}}\mid_{\mathcal{S} \times \mathcal{S}}$ and $\intsec{\mathcal{U}}\mid_{\mathcal{S} \times \mathcal{S}}$ be even lower when $S$ has VC dimension $d$?
		The following result, which is a direct consequence of Theorems~\ref{thm:10way}, \ref{thm:10way-1} and \ref{thm:10way-2}, enables us to we answer the question in the negative:

\begin{theo}
\label{thm:10way-VCdimen}
Let $1\leq d \leq n$. 
\begin{enumerate}
\item     There exists $\mathcal{S} \subseteq 2^{[n]}$ with $\mbox{VC-dim}(\mathcal{S}) \leq d$ and 
	$R^{\rightarrow} (\disj{n}\mid_{\mathcal{S}\times \mathcal{S}}) = \Omega\left( d \log \frac{n}{d} \right)$.

\item    There exists $\mathcal{S} \subseteq 2^{[n]}$ with $\mbox{VC-dim}(\mathcal{S}) \leq
      d$ and $R(\disj{n}\mid_{\mathcal{S}\times \mathcal{S}}) = \Omega\left( d \frac{\log (n/d)}{ \log \log (n/d)} \right)$.      

\item     There exists $\mathcal{S} \subseteq 2^{[n]}$ with $\mbox{VC-dim}(\mathcal{S}) \leq
      d$ and $R(\intsec{n} \mid_{\mathcal{S} \times \mathcal{S}}) = \Omega\left( d \log
        \frac{n}{d} \right)$.
\end{enumerate}
\end{theo}

The following table compares our result with the previous best known lower bound for $\disj{\mathcal{U}}\mid_{\mathcal{S} \times \mathcal{S}}$ and $\intsec{\mathcal{U}}\mid_{\mathcal{S} \times \mathcal{S}}$ among all sets $S\subset 2^{\mathcal{U}}$ of VC dimension $d$.

\begin{table}[ht]
\centering
\begin{tabular}{|c | c | c | c | c |} 
 \hline
  Problems & $R(\disj{n}\mid_{\mathcal{S}\times \mathcal{S}})$ & $R^{\rightarrow}(\disj{n}\mid_{\mathcal{S}\times \mathcal{S}})$ & $R(\intsec{n}\mid_{\mathcal{S}\times \mathcal{S}})$ & $R^{\rightarrow}(\intsec{n}\mid_{\mathcal{S}\times \mathcal{S}})$ \\ [0.5ex]
 \hline\hline 
 Previously Known & $\Omega(d)$ &   $\Omega(d\log d)$ & 
 $\Omega(d)$ & $\Omega(d\log d)$ \\ [1ex]
 & \cite{HastadW07} & \cite{DasguptaKS12} & \cite{BrodyCKWY14} & \cite{BrodyCKWY14}
 \\ [1ex]
 \hline
 This Paper & $\Omega\left(d\frac{\log (n/d)}{\log\log (n/d)}\right)$ & $\Omega\left(d\log\frac{n}{d}\right)$ & $\Omega\left(d\log\frac{n}{d}\right)$ & $\Omega\left(d\log\frac{n}{d}\right)$\\ [1ex] 
\hline
\end{tabular}
\caption{The largest communication complexity, for the functions $\disj{n}\mid_{\mathcal{S}\times \mathcal{S}}$ and $\intsec{n}\mid_{\mathcal{S}\times \mathcal{S}}$,
among all $S\subseteq 2^{[n]}$ of VC dimension $d$,
that was previously known and what we prove in this paper. Tight bounds of $\Omega\left(d\log\frac{n}{d}\right)$ for the largest $D(\disj{n}\mid_{\mathcal{S}\times \mathcal{S}})$,  
$D^{\rightarrow}(\disj{n}\mid_{\mathcal{S}\times \mathcal{S}})$, $D(\intsec{n}\mid_{\mathcal{S}\times \mathcal{S}})$ and 
$D^{\rightarrow}(\intsec{n}\mid_{\mathcal{S}\times \mathcal{S}})$, among all $S\subset 2^{[n]}$ of VC dimension $d$, follows directly from the fact that if $\mathcal{S}$ is a collection of all subsets of $[n]$ of size at most $d$ 
then $D(\disj{n}\mid_{\mathcal{S}\times \mathcal{S}}) = D(\intsec{n}\mid_{\mathcal{S}\times \mathcal{S}}) = \Omega \left( d \log \left( \frac{n}{d} \right) \right)$, see~\cite{10.5555/264772,rao_yehudayoff_2020}.}
\label{table:1}
\end{table}

\subsection*{Notations} 

We denote the set $\{1,\ldots,n\}$ by $[n]$.  
For a binary number ${\bf x}$, $\val{{\bf x}}$ denotes the decimal value of ${\bf x}$. For two vectors ${\bf x}$ and ${\bf y}$ in $\{0,1\}^n$, ${\bf x} \cap
{\bf y}=\{i \in [n]:{\bf x}_i={\bf y}_i=1\}$, and ${\bf x} \subseteq {\bf y}$ when ${\bf x}_i \leq {\bf y}_i$ for each $i \in [n]$. For a finite set $X$, $2^X$ denotes the power set of $X$. For $x,y \in \R$ with $x< y$, $[x,y]$ denotes the closed interval is the set of all real numbers that lies between $x$ and $y$

\section{One way communication complexity (Theorems~\ref{thm:10way} and~\ref{thm:10way-VCdimen}~(1))}

In this section we will prove the following result.
\begin{theo}
    For all $n \geq d$, 
      there exists 
      $X \subset \Z$ with $\size{X}=n$
      and $\cR \subseteq 2^{X}$ with $\mbox{VC-dim}(\mathcal{R}) =2d$, such that
      $$
        \cR \subseteq \left\{ X \cap\left( \bigcup\limits_{1\leq j\leq d} I_{j}\right) \, \mid \,
        \left\{I_{1}, \, \dots, \, I_{d}\right\} \in \cI
    \right\}\mbox{ and }
      	R^{\rightarrow} (\mbox{\sc Disj}_{X}\mid_{{\cal R} \times {\cal R}}) = \Omega\left( d \log \frac{n}{d} \right).
      $$
        Note that the set $\cI$ is defined in Problem~\ref{problem2}.
      \label{theo-vc-set-disjointness-oneway-1}
\end{theo}

\begin{rem}
    The above result takes care of the proofs of Theorem~\ref{thm:10way} and Theorem~\ref{thm:10way-VCdimen}~(1).
\end{rem}

\remove{
Recall that $R\left(\mbox{{\sc Disj}}_{[n]}|_{2^{[n]} \times 2^{[n]}}\right)=\Omega(n)$~\cite{KalyanasundaramS92}. As $\mbox{VC-dim}({2^{[n]}})=n$, the above theorem holds for $d=\Omega(n)$. Hence, we safely assume $d=o(n)$ in the rest of the proof. Also, without loss of generality assume that $d$ divides $n$. \textcolor{blue}{Manaswi: maybe we should mention for all $d$ in the theorem above.}
}

The \emph{hard} instance, for the proof of the above theorem, is inspired by the \emph{interval} set systems in combinatorial geometry and is constructed in Section~\ref{sec:oneway-hard}. In Section~\ref{sec:proof-oneway}, we proof Theorem~\ref{theo-vc-set-disjointness-oneway-1} by using a reduction from {\sc Augmented Indexing},
which we 
denote by \augind{\ell}. Formally the problem \augind{\ell} is defined as follows: Alice gets a string ${\bf x} 
\in \{0,1\}^{\ell}$ and Bob gets an index 
$j \in [\ell]$ and all $x_{j'}$ for $j'<j$. 
Bob reports $x_j$ as the output. \remove{$\mbox{\augind{\ell}}({\bf x},j) = x_j$.}
\remove{
\defproblem{\augind{n}}{Alice gets a string ${\bf x} \in \{0,1\}^{\ell}$ and Bob gets an index 
$j \in [n]$ and all $x_{j'}$ for $j'<j$.}{Bob reports $\mbox{\augind{n}}({\bf x},j)$ which is $x_j$.}
}

\begin{pro}\rm{(Miltersen et al.~\cite{MiltersenNSW98})}
$R^{\rightarrow}(\mbox{\augind{\ell}})=\Omega(\ell)$.
\end{pro}

\subsection{Construction of a hard instance}
\label{sec:oneway-hard}
\remove{Before proceeding further towards the proof of Theorem~\ref{theo-vc-set-disjointness-oneway-1}, we first describe $X \subset \Z$ with $\size{X}=n$ and $\cR \subseteq 2^X$ with $\mbox{VC-dim}(\mathcal{R}) =2d$.}
We construct a set $X \subset \Z$ with $\size{X}=n$ and $\cR \subseteq 2^X$ with $\mbox{VC-dim}(\cR)=2d$. Informally, $X$ is the union of the set of points present in the union of $d$ pairwise disjoint intervals, in $\Z$, each containing $\frac{n}{d}$ points. Each set in $\cR$ is the union of the set of points in the subintervals anchored either at the left or the right end point of each of the above $d$ intervals. Formally, the description of $X$ and $\cR$ are given below along with some of its properties that are desired to show Theorem~\ref{theo-vc-set-disjointness-oneway-1}.

\paragraph*{The ground set $X$:} Let $m=\frac{n}{d}-2$. Without loss of generality we can assume that $m = 2^{k}$, where $k \in \mathbb{N}$.
Let $J_{0}=\{0,\ldots,m+1\}$ be the set of $m+2$ consecutive integers that starts from the origin and ends at $m+1$. Similarly, let $J_{p}$ be the set of $m+1$ consecutive integers that starts at $p \in \Z$ and ends at $p+m +1$. 
Let $p_{1}$, $\dots$, $p_{d}$ be $d$ points in $\mathbb{Z}$ such that the sets $J_{p_1},\, \ldots,\, J_{p_d}$ are pairwise disjoint. Let the \emph{ground} set $X$ be $\bigcup\limits_{i=1}^d J_{p_i}$. Note that $X \subset \Z$ and $\size{X}=(m+2)d=n$. 

\paragraph*{The subsets of $X$ in $\cR$:} ${\cal R} \subseteq 2^X $ contains two types of sets ${\cal R}_0$ and ${\cal R}_{m+1}$, where
\begin{itemize}
\item Take any $d$ intervals $R_1, \ldots,R_{d}$ of integer lengths such that, for all $i\in [d]$, length of $R_{i}$ is at most $m+1$, $R_{i} \subseteq [p_{i}, p_{i}+m+1]$, and $R_{i}$ starts at $p_i$. Note that $R_i$ does not intersect with any $X \setminus J_{p_i}$. The set $A=\bigcup\limits_{i=1}^d \left(R_i \cap X \right)$ is an element in ${\cal R}_0$. We say that $A$ is \emph{generated} by $R_1,\ldots,R_d$.
\item Take any $d$ intervals $R'_1, \ldots,R'_d$ of integer lengths such that, for all $i \in [d]$, length of $R'_{i}$ is at most $m+1$, $R'_i \subseteq [p_{i}, p_{i}+m+1]$ and  $R'_{i}$ ends at $p_i+m+1$. Note that $R'_i$ does not intersect with any $X \setminus J_{p_i}$. The set $B=\bigcup\limits_{i=1}^d \left(R'_i\cap X\right)$  is an element in ${\cal R}_{m+1}$. We say that $B$ is \emph{generated} by $R'_1,\ldots,R'_d$.
\end{itemize}

The following claim bounds the VC dimension of $\cR$, constructed as above. 
\begin{cl}
For $X \subset \Z$ with $\size{X}=n$ and $ \cR \subset 2^X$ as described above, $\mbox{VC-dim}(\mathcal{R}) =2d$,
\end{cl}
\begin{proof}
The proof follows from the fact that any subset of of $X$ containing $2d+1$ points will contain at least three points from some $J_{p_{i}}$, where $i \in [d]$. These points in $J_{p_{i}}$ can not be shattered by the sets in $\cR$.  Also, observe that there exists $2d$ points, with two from each $J_{p_{j}}$, that can be shattered by the sets in $\cR$.
\end{proof}
Now, we give a claim about $X$ and $\cR$ constructed above that will be required for our proof of Theorem~\ref{theo-vc-set-disjointness-oneway-1}. 
\begin{cl}\label{obs:interval-oneway}
Let $A \in {\cal R}_0$ and $B \in {\cal R}_{m+1}$ be such that $A$ is generated by $R_1,\ldots,R_d$ and $B$ is generated by $R'_1,\ldots,R'_d$. Then $A$ and $B$ intersects if and only if there exists an $i \in [d]$ such  that $R_i$ intersects $R'_i$ at a point in $J_{p_i}$. 
\end{cl}
The proof of Claim~\ref{obs:interval-oneway} follows directly from our construction of $X\subset \Z$ and $\cR \subseteq 2^X$, as $J_{p_1},\, \ldots,\, J_{p_d}$ are pairwise disjoint.

\subsection{Reduction from \augind{d \log m} to  $\mbox{{\sc Disj}}_{X}\mid_{\cR \times \cR}$} 
\label{sec:proof-oneway}

Before presenting the reduction we recall the definitions of \augind{d \log m} and $\mbox{{\sc Disj}}_{X}|_{\cR \times \cR}$. In \augind{d \log m}, Alice gets ${\bf x} \in \{0,1\}^{d \log m}$ and Bob gets an index $j$ and $x_{j'}$ for each $j'<j$. The objective of Bob is to report $x_j$ as the output. In $\mbox{{\sc Disj}}_{X}|_{\cR \times \cR}$, Alice gets $A \in \cR_0$ and Bob gets $B \in \cR_{m+1}$. The objective of Bob is to determine whether 
$A \cap B = \emptyset$. Note that $X,\cR,\cR_0$ and $\cR_{m+1}$ are as discussed in the Section~\ref{sec:oneway-hard}.

Let ${\cal P}$ be an one-way protocol that solves $\mbox{{\sc Disj}}_{X}|_{\cR \times \cR}$ with $o\left( d 
\log \frac{n}{d}\right)=o(d \log m)$ bits of communication. Now, we consider the following protocol ${\cal 
P}'$ for \augind{d \log m} that has the same one way communication cost as that of $\mbox{{\sc Disj}}_{X}|_{\cR \times \cR}$. Then we will be done with the proof of Theorem~\ref{theo-vc-set-disjointness-oneway-1}.
 
 \paragraph*{Protocol ${\cal P}'$ for \augind{d \log m} problem}
 \begin{description}
 \item[Step-1] Let ${\bf x} \in \{0,1\}^{d \log m}$ be the input of Alice. Bob gets an index $j \in [d \log m]$ and bits $x_{j'}$ for each $j' < j$.
 
 \item[Step-2] Alice will form $d$ strings ${\bf a}_1,\ldots, {\bf a}_{d} \in \{0,1\}^{\log m}$ by partitioning the 
 string ${\bf x}$ into $d$ parts such that ${\bf a}_i=x_{(i-1)\log m+1}\ldots x_{i \log m} $, where $i \in [d]$. Bob first forms a string ${\bf y} \in \{0,1\}^{d\log m}$, where $y_{j'}=x_{j'}$ for each $j' < j$, $y_j = 1$, and $y_{j'}=0$ for each $j'>j$. Then Bob finds ${\bf b}_1,\ldots, {\bf b}_{d} \in \{0,1\}^{\log m}$ by partitioning the 
 string ${\bf y}$ in to $d$ parts such that ${\bf b}_i=y_{(i-1)\log m+1}\ldots y_{ i \log m} $, where $i \in [d]$.
 
 \item[Step-3] For each $i \in [d]$, let $R_i$ and $R_i'$ be the intervals that starts at $p_i$ and ends at $p_i+m+1$, respectively, where $R_i=[p_i,m+p_i-\val{{\bf a}_i}]$ and $R'_i=[p_{i}+m+1-\val{{\bf b }_i},p_i+m+1]$. Alice finds the set $A \in {\cal R}_0$ generated by $R_1,\ldots,R_d$ and Bob finds the set $B \in {\cal R}_{m+1}$ generated by $R'_1,\ldots,R'_d$, i.e., $A=\bigcup\limits_{i \in [d]}(R_i \cap X)$ and $B=\bigcup\limits_{i \in [d]}(R'_i \cap X)$.
 
 \item[Step-4] Alice and Bob solves $\disj{X}\mid_{{\cal R} \times {\cal R}}$ on inputs $A$ and $B$, and 
 report $x_j=0$ if and only if $\disj{X}\mid_{{\cal R} \times {\cal R}}(A,B)=0$. Note that $x_j$ is the output 
 of  \augind{d \log m} problem.
 \end{description}
 
 The following observation follows from the description of the protocol ${\cal P}'$ and from the construction of $X \subset \Z$ and $\cR \subseteq 2^X$. 

\begin{obs}\label{obs:correct}
Let $i^* \in [d]$ such that $j \in \{(i^*-1) \log m+1,i^* \log m\}$. Then
\begin{itemize}
\item[(i)] $R_i \cap R'_i =\emptyset$ for all $i \neq i^*$.
\item[(ii)] $R_{i^*} \cap R'_{i^*} =\emptyset$ if and only if $\val{{\bf b}_{i^*}} \leq  \val{{\bf a}_{i^*}}$.
\item[(iii)]  $\val{{\bf b}_{i^*}} \leq  \val{{\bf a}_{i^*}}$ if and only if $x_j=0$.
\end{itemize}
\end{obs}
We will use the above observation to show the correctness of the protocol ${\cal P}'$. 

First consider the case $\disj{X}\mid_{{\cal R} \times {\cal R}}(A,B)=0$.
Then, by Claim~\ref{obs:interval-oneway}, there exists an $i \in [d]$ such that $R_{i}$ and $R'_{i}$ intersects at a point in $J_{p_i}$. From Observation~\ref{obs:correct}~(i), we can say $R_{i*} \cap R'_{i^*} \neq \emptyset$.  Combining $R_{i*} \cap R'_{i^*}\neq \emptyset$ with Observations~\ref{obs:correct}~(ii) and (iii), we have $x_j=0$. Hence, $\disj{X}\mid_{{\cal R} \times {\cal R}}(A,B)=0$ implies $x_j=0$. The converse part, i.e., $x_j=0$ implies $\disj{X}\mid_{{\cal R} \times {\cal R}}(A,B)=0$, can be shown in the similar fashion.

The one-way communication complexity of protocol ${\cal P}'$ for \augind{d \log m} is the same as that of ${\cal P}$ for $\disj{X}|_{\cR \times \cR}$, that is, $o(d \log m)$. However, this is impossible as the one-way communication complexity of {\sc Augmented Indexing}, 
over $d \log m$ bits, is $\Omega(d \log m) = \Omega \left( d \log \frac{n}{d}\right)$ bits. This completes the 
proof of Theorem~\ref{theo-vc-set-disjointness-oneway-1}.

\section{Two way communication complexity (Theorems~\ref{thm:10way-1},~\ref{thm:10way-2},~\ref{thm:10way-VCdimen}(2) and ~\ref{thm:10way-VCdimen}(3))}\label{sec:vclower}
In this section, we prove the following theorems.

\begin{theo}\label{theo:main}
     For all $n \geq d$, there exists a $G \subset \Z^2$ with $\size{G}=n$ and $\cT \subseteq 2^{G}$ with $\mbox{VC-dim}(\cT)=2d$, such that 
      $$
     \cT \subseteq \left\{ G \cap\left( \bigcup\limits_{1\leq j\leq d} \ell_{j}\right) \, \mid \,
        \left\{\ell_{1}, \, \dots, \, \ell_{d}\right\} \in \cL
    \right\}\mbox{ and } 
      	R (\disj{G}\mid_{\cT \times \cT}) = \Omega\left( d \frac{\log (n/d)}{ \log \log (n/d)} \right).
      $$
      The set $\cL$ is as defined in Problem 1.
\end{theo}

\begin{theo}\label{theo:main1}
      For all $n \geq d$, there exists a $G \subset \Z^2$ with $\size{G}=n$ and $\cT \subseteq 2^{G}$ with $\mbox{VC-dim}(\cT)=2d$, such that 
      $$
     \cT \subseteq \left\{ G \cap\left( \bigcup\limits_{1\leq j\leq d} \ell_{j}\right) \, \mid \,
        \left\{\ell_{1}, \, \dots, \, \ell_{d}\right\} \in \cL
    \right\}\mbox{ and } 
      	R (\intsec{G}\mid_{\cT \times \cT}) = \Omega\left( d \log \frac{n}{d}\right).
      $$
      The set $\cL$ is as defined in problem 1.
\end{theo}

\begin{rem}
Theorem~\ref{theo:main} takes care of Theorem~\ref{thm:10way-1} and~\ref{thm:10way-VCdimen}(2). Theorem~\ref{theo:main1} takes care of Theorem~\ref{thm:10way-2} and~\ref{thm:10way-VCdimen}(3).
\end{rem}

Note that the same set system will be used for the proofs of the above theorems. 
The \emph{hard} instance, for the proof of the above theorems, is inspired by \emph{point line incidence} set systems in computational geometry and is constructed in Section~\ref{sec:setsystem}. We prove Theorems~\ref{theo:main} and~\ref{theo:main1} in Sections~\ref{sec:twoway-disj} and~\ref{sec:twoway-int}, respectively, using reductions.

\subsection{The hard instance for the proofs of Theorems~\ref{theo:main} and~\ref{theo:main1}} 
\label{sec:setsystem}
In this subsection, we give the description of $G \subset \Z^2$ with $\size{G}=n$ and $\cT \subseteq 2^G$, with $\mbox {VC-Dim}(\cT)=2d$. The same $G$ and $\cT$ will be our  \emph{hard} instance for the proofs of  Theorems~\ref{theo:main} and~\ref{theo:main1}. In this subsection, without loss of generality, we can assume that $d$ divides $n$ and $n/d$ is a perfect square. 

Informally, $G$ is the set of points present in the union of $d$ many pairwise disjoint square grids each containing $\frac{n}{d}$ points and the grids are taken in such a way that any straight line of non-negative can intersects with at most one grid. Also, each set in $\cT$ is the union of the set of points present in $d$ many lines of non-negative slope such that one line intersects with exactly one grid. Moreover, all of the $d$ lines have slopes   either zero or positive. Formally, the description of $G$ and $\cT$ are given below along with some of its properties that are desired to show Theorems~\ref{theo:main} and~\ref{theo:main1}.
\paragraph*{The ground set $G$:}
Let $m=\sqrt{\frac{n}{d}}$, and $G_{(0,0)} = \left\{(x,y)\in \Z^2:0 \leq x,y \leq m-1 \right\}$ be the grid of size $m \times m$ anchored at the origin $(0,0)$. For any $p, q \in \Z$, the $m\times m$ grid anchored at $(p,q)$ will be denoted by $G_{(p,q)}$, i.e., $G_{(p,q)} = \left\{(i+p,j+q) : (i,j) \in G_{(0,0)} \right\}$. For $d \in \N$, consider $G_{(p_1,q_1)}, \ldots, G_{(p_{d},q_{d})}$ satisfying the following property:

\begin{center}
    \underline{{\sc Property}} {\em For any $i, \, j\in [d]$, with $i\neq j$, let $L_1$ and $L_2$ be lines of non-negative slopes that pass through at least two points of $G_{(p_i,q_i)}$ and $G_{(p_j,q_j)}$, respectively. Then $L_1$ and $L_2$ does not intersect at any point \emph{inside} $\bigcup_{\ell=1}^{d} G_{(p_\ell,q_\ell)}$.}
\end{center}

Observe that there exists $G_{(p_1,q_1)}, \ldots, G_{(p_{d},q_{d})}$ satisfying 
{\sc Property}. We will take $G = \bigcup_{\ell=1}^{d} G_{(p_\ell,q_\ell)}$ as the ground set. Without loss of generality, we can assume that $(p_1,q_1)=(0,0)$. Note that $G \subset \Z^2$ and $\size{G}=dm^2=n$.
 \remove{All further arguments in this section, will be on $G$ satisfying {\sc Property}.}
\paragraph*{The subsets of $G$ in $\cT$:}
$\cT$ contains two types of subsets $\cT_{1}$ and $\cT_{2}$ of $G$, and they are generated by the following ways:
\begin{itemize}
\item 
    Take any $d$ lines $L_1,\ldots,L_d$ of non negative slope such that, $\forall i \in [d]$, $L_{i}$ passes through $(p_{i}, q_{i}) \in G_{(p_i,q_i)}$ and (at least) another point in $G_{(p_i,q_i)}$. Note that  $L_i$ does not contain any point from $G \setminus G_{(p_i,q_i)}$. The set $A=\bigcup_{i=1}^d \left( L_i \cap G_{(p_i,q_i)} \right)$ is in $\cT_1$, and we say $A$ is \emph{generated} by the lines $L_1,\ldots,L_d$.

\item 
    Take any $d$ vertical lines $L'_1,\ldots,L'_d$ such that, $\forall i \in [d]$, $L_{i}'$ contains at least one point from $G_{(p_i,q_i)}$. Note that  $L'_i$ does not contain any point from $G \setminus G_{(p_i,q_i)}$. The set $B=\bigcup_{i=1}^d (L'_i \cap G_{(p_i,q_i)})$ is in $\cT_2,$ and we say $B$ is generated by the lines $L'_1,\ldots,L'_d$. 
\end{itemize} 
The following claim bounds the VC dimension of $\cT$, which as described above.
\begin{cl}
For $G \subset \Z^2$ and $\cT \subseteq 2^G$ as described above, ${\mbox{VC-dim}}(\cT)=2d$.
\end{cl}
\begin{proof}
The proof follows from the fact that any subset of $X$ containing $2d+1$ points will contain at least three points from some $G_{(p_j,q_j)}, j \in [d]$. These points in $G_{(p_j,q_j)}$ can not be shattered by the sets in $\cT$.  Also, observe that there exists $2d$ points two from each $G_{(p_j,q_j)}$ that can be shattered by the sets in $\cT$.
\end{proof}

Now, we give two claims about $G$ and $\cT$, constructed above, that follow directly from our construction of $G \subset \Z^2$ and $\cT \subseteq 2^{G}$.

\begin{cl}\label{cl:inter1}
Let $A \in \cT_1$ and $B \in \cT_2$ such that $A$ is generated by lines $L_1,\ldots,L_d$ and $A$ is generated by lines $L'_1,\ldots,L'_d$. Then $A$ and $B$ intersect if and only if there exists $i \in [d]$ such that $L_i$ and $L_i'$ intersect at a point in $G_{(p_i,q_i)}$.  
\end{cl}

\begin{cl}\label{cl:inter}
Let $A \in \cT_1$ and $B \in \cT_2$ such that $A$ is generated by lines $L_1,\ldots,L_d$ and $B$ is generated by lines $L'_1,\ldots,L'_d$. Also let $\size{A \cap B}=d$. Then for each $i \in [d]$, $L_i$ and $L_i'$ intersect at a point in $G_{(p_i,q_i)}$. Moreover, $A$ ($B$) can be determined if we know $B$ ($A$) and $A \cap B$.
\end{cl}
The above claims will be used in the proofs of Theorems~\ref{theo:main} and~\ref{theo:main1}.

\subsection{Proof of Theorem~\ref{theo:main}}\label{sec:twoway-disj}

\remove{We assume that $d=o(n)$, as otherwise the theorem follows from the fact that $R\left(\mbox{{\sc Disj}}_{[n]}|_{2^{[n]} \times 2^{[n]}}\right)=\Omega(n)$~\cite{KalyanasundaramS92}. With out loss of generality, we also assume that $d$ divides $n$ and, more over, $n/d$ is a perfect square.}

Let us consider a problem in communication complexity denoted by $\mbox{{\sc Or-Disj}}_{\{0,1\}^{\ell}}^t$ that will be used in our proof. In $\mbox{{\sc Or-Disj}}_{\{0,1\}^{\ell}}^t$, Alice gets $t$ strings ${\bf x}_1,\ldots,{\bf x}_t \in \{0,1\}^{\ell}$ and Bob also gets $t$ strings ${\bf y}_1,\ldots,{\bf y}_t \in \{0,1\}^{\ell}$. The objective is to compute 
$\bigvee \limits_{i=1}^t \mbox{{\sc Disj}}_{\{0,1\}^{\ell}}({\bf x}_i,{\bf y}_i)$. Note that $\disj{\{0,1\}^{\ell}}({\bf x}_i,{\bf y}_i)$ is a binary variable that takes value $1$ if and only if ${\bf x}_i \cap {\bf y}_i =\emptyset $.

\begin{pro}[Jayram et al.~\cite{JayramKS03}]\label{lem:disj1}
$R\left(\mbox{{\sc Or-Disj}}_{\{0,1\}^{\ell}}^t \right)=\Omega(\ell t)$.
\end{pro}

Note that Proposition~\ref{lem:disj1} directly implies the following result. 

\begin{pro}\label{lem:disj}
    $R\left(\mbox{{\sc Or-Disj}}_{\{0,1\}^{\ell}}^t \mid_{S_{\ell} \times S_{\ell}} \right)=\Omega(\ell t)$, where ${S}_{\ell} = \{0,1\}^{\ell} \setminus \{0^{\ell}\}$.
\end{pro}

Let $k \in \mathbb{N}$ be the largest integer such that first $k$ consecutive primes $p_{1}, \, \dots, \, p_{k}$ satisfy the following inequalty:
\begin{equation}\label{eqn:product-k-primes}
    \Pi_{i=1}^{k} p_{i} \leq \sqrt{\frac{n}{d}}.
\end{equation}
Using the fact that $\Pi_{i=1}^{k} p_{i} = e^{(1+o(1))k\log k}$, we get $k = \Theta\left( \frac{\log (n/d)}{\log \log (n/d)} \right)$.

We prove the theorem by a reduction from $\mbox{{\sc Or-Disj}}_{\{0,1\}^{k}}^{d}\mid_{S_{k} \times S_{k}}$ to $\mbox{{\sc Disj}}_{G}
\mid_{\cT \times \cT}$, where 
$$
    S_{k} := \{0,1\}^{k} \setminus \{0^{k}\}.
$$
Note that $G \subset \Z^2$ with $\size{G}=n$, and $\cT \subseteq 2^{G}$, with $\mbox{VC-dim}(\cT)=2d$, are the same as that we constructed in 
Section~\ref{sec:setsystem}. 
To reach a contradiction, assume that there exists a two way protocol $\cP$ that solves $\mbox{{\sc Disj}}_{G}~
\mid_{\cT \times \cT}$ with communication cost of $o\left( d \frac{\log m}{\log \log m}\right)=o\left( d \frac{\log (n/d)}{ \log \log (n/d)} \right)$ bits.  Now, we give  protocol $\cP'$ that solves $\mbox{{\sc Or-Disj}}_{\{0,1\}^{k}}^{d} \mid_{S_{k} \times S_{k}}$, as described below.

\paragraph*{Protocol $\cP'$ for $\mbox{{\sc Or-Disj}}_{\{0,1\}^{k}}^{d} \mid_{S_{k} \times S_{k}}$}
\begin{description}

    \item[Step-1] 
        Let $A=({\bf x}_1,\ldots, {\bf x}_d) \in \left[S_{k} \right]^{d}$~\footnote{For a set $W$, $[W]^d = W \times \ldots \times W$ ($d$ times).} and $B=({\bf y}_1,\ldots, {\bf y}_d) \in \left[ S_{k} \right]^{d}$ be the inputs of Alice and Bob for $\mbox{{\sc Or-Disj}}_{\{0,1\}^{ k }}^{d} \mid_{S_{k} \times S_{k}}$. Recall that $S_{k} = \left\{0,1\right\}^{k}\setminus \{0^{k}\}$.
        Bob finds $ \bar{ B}=(\bar{{\bf y}}_1,\ldots, \bar{{\bf y}}_d) \in \left[\{0,1\}^{k} \right]^d$, where $\bar{{\bf y}}_i$ is obtained by complementing each bit of ${\bf y}_i$.

    \item[Step-2] 
        Both Alice and Bob privately determine first $k$ prime numbers $p_{1},\ldots,p_{k}$ without any communication. 
        

 \item[Step-3] Let $\Phi:\{0,1\}^{k} \to \{0,1\}^{\left\lceil \log \left( \sqrt{\frac{n}{d}}\right)\right\rceil}$ be the function such that $\phi({\bf x})$ is the bit representation of the number $\prod_{i=1}^{k} p_{i}^{x_{i}}$, where $\mathbf{x} = (x_{1},\dots, x_{k}) \in \{0,1\}^{k}$. Alice finds $A' = (\mathbf{a}_1,\dots,\mathbf{a}_{d}) \in \left[\{0,1\}^{\left\lceil \log \left( \sqrt{\frac{n}{d}}\right)\right\rceil}\right]^{d}$ and Bob finds $B' = (\mathbf{b_1},\ldots,\mathbf{b_1}\in \left[ \{0,1\}^{\left\lceil \log \left( \sqrt{\frac{n}{d}}\right)\right\rceil}\right]^{d}$ privately without any communication, where ${\bf a_i}=\phi(\bf x_i)$ and ${\bf b_i}=\phi(\bar{{\bf y}}_i)$ for each $i \in [d]$.

\item[Step-4] For each $i \in [d]$, let $L_i$ and $L_i'$ be the lines having equation $y-q_i=\frac{\val{{\bf a}_i}-1}{\val{{\bf a}_i}}(x-p_i)$ and $x-p_i=\val{{\bf b}_i}$ respectively. Alice finds $A'' \in \cT$ that is generated by the lines $L_1,\ldots,L_d$, and Bob finds $B'' \in \cT$ which is generated by the lines $L'_1,\ldots,L'_d$, i.e., $A'' =\bigcup\limits_{i \in [d]}(L_i \cap G_{(p_i,q_i)})$ and $B'' =\bigcup\limits_{i \in [d]}(L'_i \cap G_{(p_i,q_i)})$.

\item[Step-5] Then Alice and Bob solve $\disj{G} \mid_{\cT \times \cT} (A'',B'')$, and report $
\bigvee \limits_{i=1}^{d} \mbox{{\sc Disj}}_{{\{0,1\}}^{k}}({\bf x}_i, {\bf y}_i)=1$ if and only if $
\disj{G} \mid_{\cT \times \cT}(A'',B'')=0$.
\end{description}

Now we argue for the correctness of the protocol $\cP'$. Let $\disj{G} \mid_{\cT \times \cT}(A'',B'')=0$, that is, $A'' \cap B'' \neq \emptyset $. By Claim~\ref{cl:inter1} and from the description of $\cP'$, there exists $i \in [d]$ such that the lines $L_i:y-q_i=\frac{\val{{\bf a}_i}-1}{\val{{\bf a}_i}}(x-p_i)$ and $L_i':x-p_i=\val{\bf{b_i}}$ intersect at a point in $G_{(p_i,q_i)}$, that is, the lines $y=\frac{\val{{\bf a}_i}-1}{\val{{\bf a}_i}}x$ and $ x=\val{{\bf b}_i}$ intersect at a point in $G_{(0,0)}$. Now, we can say that, there exists $i \in [d]$ such that $\val{{\bf a}_i}$ divides $\val{{\bf b_i}}$, equivalently, $\phi({{\bf x}_i})$ divides $\phi({\bar{{\bf y}}_{i}})$. This implies ${{\bf x}_i}$ is a subset of 
$\bar{{\bf y}}_{i}$ ( or ${{\bf x}_i} \cap {{\bf y_i}}=\emptyset$) for some $i \in [d]$. Hence,  $\bigvee \limits_{i=1}^{d} \mbox{{\sc Disj}}_{{\{0,1\}}^{ k}}({\bf x}_i, {\bf y}_i)=1$. The converse part, that is, $\bigvee \limits_{i=1}^{d} \mbox{{\sc Disj}}_{{\{0,1\}}^{ k}}({\bf x}_i, {\bf y}_i)=1$ implies $\disj{G} \mid_{\cT \times \cT}(A'',B'')=0$ can be shown in the similar fashion.

Observe that the communication cost of protocol $\cP'$ for $\mbox{{\sc Or-Disj}}_{\{0,1\}^{ k }}^{d} \mid_{S\times S}$ is same as that of protocol $\cP$ for $\disj{G} \mid_{\cT \times \cT}$, which is $o\left( d\frac{ \log m}{\log \log m}\right)= o\left( d\frac{ \log (n/d)}{\log \log (n/d)}\right) =o(dk)$ as $m = \sqrt{\frac{n}{d}}$ and $k = \Theta\left( \frac{\log (n/d)}{\log \log (n/d)} \right)$. This contradicts Proposition~\ref{lem:disj} which says that $R\left( \mbox{{\sc Or-Disj}}_{\{0,1\}^{ k }}^{d} \mid_{S\times S}\right) = \Omega (dk)$.


\subsection{Proof of Theorem~\ref{theo:main1}}\label{sec:twoway-int}
\remove{We assume that $d=o(n)$, as otherwise the theorem follows from the fact that $R\left(\mbox{{\sc Disj}}_{[n]}|_{2^{[n]} \times 2^{[n]}}\right)=\Omega(n)$~\cite{KalyanasundaramS92}.} With out loss of generality, we also assume that $d$ divides $n$ and, more over, $n/d$ is a perfect square.

First, consider the problem $\mbox{{\sc Learn}}_{G} \mid_{\cT \times \cT}$, where the objective of Alice and Bob is to learn 
each other's set. Note that $G \subset \Z^2$ with $\size{G}=n$ and $\cT \subseteq 2^{G}$ with $\mbox{VC-Dim}(\cT)=2d
$ are same as that constructed in Section~\ref{sec:setsystem}. In $\mbox{{\sc Learn}}_{G} \mid_{\cT \times \cT}$, Alice and Bob 
get two sets $A$ and $B$, respectively, from $\cT$ with a promise $\size{A \cap B}=d$. The objective of Alice (Bob) is to 
learn $B$~($A$). Observe that $R(\mbox{{\sc Learn}}_{G} \mid_{\cT \times \cT})=\Omega(d \log n)$ as there are $\Omega(m^d)=
\Omega\left( \left(\sqrt{\frac{n}{d}} \right)^d\right)$ many candidate sets for the inputs of Alice and Bob. We 
prove the theorem by a reduction from $\mbox{\mbox{{\sc Learn}}}_{G} \mid_{\cT \times \cT}$ to $\intsec{G}\mid_{\cT \times \cT}$.

 Let by contradiction consider a protocol ${\cal P}$ that solves $\intsec{G}\mid_{\cT \times \cT}$ by using $o(d \log 
 n)$ bits of communication. To solve $\mbox{{\sc Learn}}_{G} \mid_{\cT \times \cT}$, Alice and Bob first run a protocol ${\cal 
 P}$ and finds $A \cap B$. Now by Claim~\ref{cl:inter1}, it is possible for Alice (Bob) to determine $B$ ($A$) by 
 combining $A$ ($B$) along with $A \cap B$, with out any communication with Bob (Alice). Now, we have a protocol ${\cal 
 P}'$ that solves $\mbox{{\sc Learn}}_{G} \mid_{\cT \times \cT}$ with $o(d \log n)$ bits of communication. However, this is impossible as $R(\mbox{{\sc Learn}}_{G} \mid_{\cT \times \cT})=\Omega(d \log n)$. Hence, we are done with the proof of Theorem~
 \ref{theo:main1}.

\section{Conclusion and Discussion}
\label{sec:conclude}

In this paper, we studied  $\disj{n}\mid_{\mathcal{S} \times \mathcal{S}}$ and $\intsec{n}\mid_{\mathcal{S} \times 
\mathcal{S}}$ when $\mathcal{S}$ is a subset of $2^{[n]}$ and $\mbox{VC-dim}(\mathcal{S}) \leq d$.
One of the main contributions of our work is the result (Theorem~\ref{thm:10way-VCdimen}) showing that unlike in the case of $\spdisj{n}{d}$ and $\spdisj{n}{d}$ functions, there is no
separation between randomized and deterministic communication
complexity of $\disj{n}\mid_{\mathcal{S} \times \mathcal{S}}$ and
$\intsec{n}\mid_{\mathcal{S} \times \mathcal{S}}$ functions when $\mbox{VC-dim}(S) \leq d$. Note that we have settled both the one-way and two-way (randomized) communication complexities of $\intsec{n}\mid_{\cS \times \cS}$ when $\mbox{VC-dim}(\mathcal{S}) \leq d$~(Theorem~\ref{thm:10way-VCdimen} (1) and (3)). In the context of $\disj{n}\mid_{\mathcal{S} \times \mathcal{S}}$, we have settled the one-way (randomized) communication complexity. The two-way communication complexity for $\disj{n}\mid_{\mathcal{S} \times \mathcal{S}}$ is tight up to factor $\log \log \frac{n}{d}$~(See Theorem~\ref{thm:10way-VCdimen} (2)). However, we believe that the factor of $\log \log \frac{n}{d}$ should not be present in the statement of Theorem~\ref{thm:10way-VCdimen} (2).
 
 \begin{conj}
 \label{conj1}
 There exists $\mathcal{S} \subseteq 2^{[n]}$ with $\mbox{VC-dim}(\mathcal{S}) \leq
      d$ and $R(\disj{n}\mid_{\mathcal{S}\times \mathcal{S}}) = \Omega\left( d  \log \frac{n}{d} \right)$.      
 \end{conj}
 
 Recall $G \subset \Z^2$ with $\size{G}=n$ and $\cT \subseteq 2^{G}$ with $\mbox{VC-Dim}(\cT)=2d$ construction 
 from Section~\ref{sec:setsystem}, that served as the hard instance for the proof of Theorem~\ref{theo:main} and Theorem~\ref{theo:main1}.  
 The same $G$ and $\cT$ can not be the hard instance for the proof of Conjecture~\ref{conj1}
because of the following result.
 \begin{theo}\label{theo:conclude}
 Let us consider $G \subset \Z^2$ with $\size{G}=n$ and $\cT \subseteq 2^{G}$ with $\mbox{VC-Dim}(\cT)=2d$ as defined in Section~\ref{sec:setsystem}. Also, recall the definition of  $\cT_1$ and $\cT_2$. There exists a randomized communication protocol that can, $\forall A \in \cT_{1}$ and $\forall B \in \cT_{2}$, can compute  
 $\disj{G}\mid_{\cT \times \cT}(A, B)$, with probability at least $2/3$, and uses $O\left(\frac{d \log d \log \frac{n}{d}}{ \log \log \frac{n}{d}} \cdot \log \log \log \frac{n}{d}\right)$ bits of communication.
 \end{theo}
 We use the following observation to prove the above theorem.
 \begin{obs}\label{obs:div}
Let us consider the communication problem $\mbox{{\sc Gcd}}_k(a,b)$, where Alice and Bob get $a$ and $b$ respectively from $\{1,\ldots,k\}$, and the objective is for both the players to compute ${\rm gcd}(a,d)$. Then there exists a randomized protocol, with success probability at least $1-\delta$, for $ \mbox{{\sc Gcd}}_{k}$ that uses $O\left( \frac{\log k}{\log \log k}\cdot \log \log \log k \cdot \log \frac{1}{\delta}\right)$ bits of communication.
 \end{obs}
 \begin{proof}
 We will give a protocol $P$ for the case when $\delta=1/3$ that uses $O\left(\frac{\log k}{\log \log k}\cdot \log \log \log k \right)$ bits of communication. By repeating $O\left(\log \frac{1}{\delta} \right)$ times protocol $\cP$ and reporting the majority of the outcomes as the output, we will get the correct answer with probability at least $1-\delta$.
 Both Alice and Bob generate all the prime numbers $p_{1},\, \dots, \, p_{t}$ between $1$ and $k$. From the Prime Number Theorem, we known that $t = \Theta\left( \frac{k}{\log k}\right)$. Alice and Bob separately, 
 construct the sets $S_{a}$ and $S_{b}$ that contain the prime numbers that divides $a$ and $b$ respectively. Note that $\size{S_{a}}$ and $\size{S_{b}}$ is bounded by $O\left(\frac{\log k}{\log \log k}\right)$.\footnote{The product of first $t$ prime numbers is $e^{(1+o(1))t\log t}$.} Alice and Bob compute $S_{a} \cap S_{b}$ by solving \emph{Sparse Set Intersection} problem on input $S_{ a}$ and $S_{ b}$ using $O\left( \frac{\log k}{\log  \log k}\right)$ bits of communication~\cite{BrodyCKWY14}. For $p \in S_{a}\cap S_{b}$, let $\alpha_{p,a}$ and $\alpha_{p,b}$ denote the exponent of $p$ in $a$ and $b$, respectively. Observe that 
 $$ 
        {\rm gcd}(a,b) = \displaystyle\prod_{p \in S_{a}\cap S_{b}} p^{\min\{\alpha_{p,a}, \alpha_{p,b}\}}.
$$      
  For each $p \in S_a$, Alice sends $\alpha_{p,a}$ to Bob.  Number of bits of communication required to send the exponents of all the primes in $S_{a} \cap S_{b}$, is 
  \begin{align*}
        \size{S_{a} \cap S_{b}}+ \sum\limits_{p \in S_{a} \cap S_{b}} \log (\alpha_{p,a})
        &\leq O\left( \frac{\log k}{\log \log k} \right)
        + |S_{a} \cap S_{b}| \log \left( \frac{\sum\limits_{p \in S_{a} \cap S_{b}} \alpha_{p,a}}{\size{S_{a} \cap S_{b}}}\right)\\
        &\leq O\left( \frac{\log k}{\log \log k} \right)
        + |S_{a} \cap S_{b}| \log \left( \frac{\log k}{\size{S_{a} \cap S_{b}}}\right)\\
        &\leq O\left(\frac{\log k}{\log \log k}\cdot \log \log \log k\right)
    \end{align*}  
    In the above inequalities, we used the facts that $\size{S_{a} \cap S_{b}} = O\left( \frac{\log k}{\log \log k}\right)$, $\sum\limits_{p \in S_{a} \cap S_{b}} \alpha_{p,a} \leq \log k$ and $\log x$ is a concave function. 
  After getting the exponents $\alpha_{p,a}$ of the primes $p \in S_{a} \cap S_{b}$ from Alice, Bob also sends the
  exponents $\alpha_{p,b}$ of the primes $p \in S_{a} \cap S_{b}$ to Alice using $O\left( \frac{\log k}{\log \log k} \log \log \log k \right)$ bits of communication to Alice. Since both Alice and Bob now know the set $S_{a} \cap S_{b}$, and the exponents $\alpha_{p, a}$ and $\alpha_{p,b}$ for all $p \in S_{a}\cap S_{b}$, both of them can compute ${\rm gcd}(a,b)$. Total number of bits communicated in this protocol is $O\left( \frac{\log k}{\log \log k} \log \log \log k \right)$.
 \end{proof}
 
 We will now give the proof of Theorem~\ref{theo:conclude}.
 
\begin{proof}[Proof of the Theorem~\ref{theo:conclude}]
Consider the case when $d=1$. From the description of $G$ and $\cT$ in Section~\ref{sec:setsystem}, we can say that $G=G_{(0,0)}$, where $G_{(0,0)}=\{(x,y) \in \Z^2: 0\leq x,y \leq \sqrt{n}\}$~\footnote{With out loss of generality assume that $\sqrt{n}$ is an integer}. Moreover, each set in $\cT_1$ is a set of points present in a straight line of non-negative slope that passes through two points of $G_{(0,0)}$ with one point being $(0,0)$ and each set in $\cT_2$ is a set of points present in a vertical straight line that passes through exactly $\sqrt{n}$ many grid points. 
Keeping Claims~\ref{cl:inter1} and~\ref{cl:inter} in mind, we will be done if we can show the existence of a randomized communication protocol for computing the function $\disj{G} \mid_{\cT \times \cT}$, with probability of success at least $1-\delta$ and number of bits communicated by the protocol being bounded by $O \left( \frac{\log n}{\log \log n}\cdot \log \log \log n \cdot \log \frac{1}{\delta}\right)$, for the special case when $d=1$. This is because for general $d$, we will be solving $d$ instances of the above problem, with the number of points in each grid being $\frac{n}{d}$ \footnote{Recall that we have assumed, without loss of generality, that $d$ divides $n$.}
and setting $\delta = \frac{1}{3d}$ for each of the $d$ instances.

\paragraph{Protocol for $d=1$.} Alice and Bob get $A$ and $B$ from $\cT_1$ and $\cT_2$, respectively. Let $A$ is generated by the straight line $L_A$ and $B$ is generated by $L_B$, where $L_A$ is a straight line with non-negative slope and $L_B$ is a vertical line. If $L_A$ is a horizontal one : Alice just sends this information to Bob and then both report that $A \cap B \neq \emptyset$.
   If $L_A$ is a vertical line : Alice sends this information to Bob and he reports $A\cap B \neq \emptyset$ if and only if $L_B$ passes through origin. \remove{If $L_B$ passes through origin : Bob sends this information to Alice and then both decide that $A \cap B \neq \emptyset$.
 n.} Now assume that $L_A$ is neither a horizontal nor a vertical line. Let the equation of $L_A$ be $y=\frac{p}{q}x$, where $1 \leq p,q \leq \sqrt{n}$, and $p$ and $q$ are relatively prime to each other. Also, let equation of Bob's line $L_B$ be $x=r$, where $0 \leq r \leq \sqrt{n}$. Observe that $A \cap B \neq \emptyset$ if and only if $L_A$ and $L_B$ intersects at a point of $G_{(0,0)}$. Moreover, $L_A$ and $L_B$ intersects at a grid point if and only if $q$ divides $r$ and $1\leq \frac{pr}{q} \leq \sqrt{n}$. So, Alice and Bob run the communication protocol for $\mbox{{\sc Gcd}}_{\sqrt{n}}(q,r)$ to decide whether $q = {\rm gcd}(q,r)$. If $q = {\rm gcd}(q,r)$ and $1\leq \frac{pr}{q} \leq \sqrt{n}$ (again Alice and Bob can decide this using $O(1)$ bits of communications) then $A\cap B \neq \emptyset$, otherwise $A\cap B = \emptyset$. Alice and Bob can decide if $q = {\rm gcd}(q,r)$ and $1\leq \frac{pr}{q} \leq \sqrt{n}$ using $O(1)$ bits of communication.

The communication cost of our protocol is dominated by the communication complexity of $\mbox{{\sc Gcd}}_{\sqrt{n}}(q,r)$, which is equal to $O\left(\frac{\log n}{\log \log n} \log \log \log n \log \frac{1}{\delta}\right)$ by Observation~\ref{obs:div}.
\end{proof}

\section*{Acknowledgments}

Anup Bhattacharya is supported by SERB-National Post Doctoral 
Fellowship, India. Arijit Ghosh is supported by Ramanujan 
Fellowship (No. SB/S2/RJN-064/2015), India. The authors would like to thank Sudeshna Kolay and Arijit Bishnu for the many discussions in the early stages of this work.

\appendix

\section{VC dimension, and Problems~\ref{problem1} and \ref{problem2}}
\label{appendix-VCD}

\paragraph{VC dimension, and collection of $d$ lines.}
Let $G \subset \Z^2$ be a set of $n$ points in $\Z^2$. Observe, that the communication functions $\disj{G}\mid_{\cL \times \cL}$ (defined in Problem~\ref{problem1}) and $\disj{G}\mid_{\cG \times \cG}$,
where 
$$
    \cG = \left\{ G \cap\left( \bigcup\limits_{1\leq j\leq d} \ell_{j}\right) \, \mid \,
        \left\{\ell_{1}, \, \dots, \, \ell_{d}\right\} \in \cL
    \right\},
$$
are equivalent problems. 
Note that the set $\cL$ is defined in Problem~\ref{problem1}. Using standard geometric arguments, see~\cite[Chap.~10]{matousek2013lectures} and \cite[Chap.~5]{har2011geometric}, we can show that $\mbox{VC-dim}(\cG) = 2d$.

\paragraph{VC dimension, and collection of $d$ intervals.} Let $X \subset \Z$ be a set of $n$ points in $\Z$. Observe, that the communication functions $\disj{X}\mid_{\cI \times \cI}$ (defined in Problem~\ref{problem2}) and $\disj{X}\mid_{\mathcal{F} \times \mathcal{F}}$,
where 
$$
    \mathcal{F} = \left\{ X \cap\left( \bigcup\limits_{1\leq j\leq d} I_{j}\right) \, \mid \,
        \left\{ I_{1}, \, \dots, \, I_{d}\right\} \in \cI
    \right\},
$$
are equivalent problems. 
Note that the set $\cI$ is defined in Problem~\ref{problem2}. Using standard geometric arguments, as in the above case, we can show that $\mbox{VC-dim}(\mathcal{F}) = 2d$.

\phantomsection
\bibliographystyle{alpha}
\addcontentsline{toc}{section}{Bibliography}
\bibliography{CommVCdim}

\end{document}